\documentstyle[12pt]{article}
\def\be{\begin{equation}}
\def\ee{\end{equation}}
\def\beq{\begin{equation}}
\def\eeq{\end{equation}}
\def\bea{\begin{eqnarray}}
\def\eea{\end{eqnarray}}
\def\bq{\begin{quote}}
\def\eq{\end{quote}}

\def\IJMP{{\it Int.J.Mod.Phys.} }

\def\NP{{\it Nucl.Phys.} }
\def\PL{{\it Phys.Lett.} }
\def\PR{{\it Phys.Rev.} }
\def\PRL{{\it Phys.Rev.Lett.} }
\def\PRTS{{\it Physics Reports} }

\def\gappeq{\mathrel{\rlap {\raise.5ex\hbox{$>$}}
{\lower.5ex\hbox{$\sim$}}}}

\def\lappeq{\mathrel{\rlap{\raise.5ex\hbox{$<$}}
{\lower.5ex\hbox{$\sim$}}}}

\begin{document}
\pagestyle{empty}
\begin{flushright}
CERN-TH/96-77 \\
ROME1 prep. 1143/96
\end{flushright}
\vspace*{5mm}

\begin{center}
{\bf FAILURE OF LOCAL DUALITY IN INCLUSIVE}
\centerline{\bf  NON-LEPTONIC HEAVY FLAVOUR DECAYS} \\
\vspace*{0.5cm}
{\bf G.~Altarelli} \\
\vspace*{0.2cm}
Theoretical Physics Division, CERN,
CH-1211 Geneva 23 and \\
Dipartimento di Fisica, Terza Universit\`{a} di Roma, Roma \\
\vspace{0.5cm}
{\bf G.~Martinelli, S.~Petrarca and F.~Rapuano} \\
\vspace{0.2cm}
Dip. di Fisica dell'Universit\`{a} {\it La Sapienza} and \\
INFN, Sez. di Roma I \\
P.le A. Moro 2, 00185 Roma, Italy \\
\vspace*{2cm}
\noindent
{\bf ABSTRACT} \\ \end{center}
\vspace*{5mm}
We argue that there is  strong  experimental evidence in the data of $b$-
and $c$-decays that the pattern of power suppressed corrections predicted by
the short distance expansion, the heavy   quark  effective theory and  the 
assumption of local duality is not  correct for the  non-leptonic inclusive
widths.  The data indicate instead  the presence of $1/m$ corrections that
should be absent in the above theoretical framework. These corrections  can
be  simply described by replacing the heavy quark mass by the mass of the 
decaying  hadron in the $m^5$ factor in front of all the non-leptonic
widths.
\vspace {0.5cm}
\begin{flushleft}
Work 
supported in part 
by the HCM Programme of the EC under contract ERBCHRXCT930132.
\end{flushleft}
\vspace{2cm}

\begin{flushleft} CERN-TH/96-77 \\
March 1996
\end{flushleft}
\vfill\eject

\setcounter{page}{1}
\pagestyle{plain}

\vskip 12pt
\noindent
{\bf 1. Introduction}

Since the discovery of charm all attempts of constructing a satisfactory
theory of heavy flavour inclusive decay properties (lifetimes and
semileptonic branching ratios) have met considerable difficulties
\cite{bigi}. With the advent of beauty it was hoped that the substantially
increased mass of these new states would finally lead to an understanding of
their inclusive decays in terms of some adequately improved form of the QCD
parton model. But even for beauty decays, with the steady progress of
experimental information and a lot of accumulated theoretical insight, a
number of problems remains unsolved  \cite{bigi,Neub}. The main examples are
the experimental value of the average semileptonic (SL) beauty meson
branching ratio which appears to be somewhat smaller than the theoretical
predictions and the observed difference of the lifetimes of the $\Lambda_b$
baryon and of the $B$ mesons which is larger than expected. This situation
is especially deceiving in that an appealing theoretical framework has been
developed \cite{buv} for power corrections in terms of a short distance
operator expansion and the formalism of the heavy quark effective theory
\cite{neube}. The result of this approach is puzzling because it predicts
that all corrections to the leading QCD improved parton terms appear at the
order $1/m^2$ and beyond, where $m$ is the heavy-quark mass, while the
experimental findings suggest much larger corrections. However the above
method relies on the use of the operator expansion in the timelike region,
namely on the physical cut, so that, in principle, some smearing should be
applied, in the spirit of ref.~\cite{pqw},  in order to avoid the infrared
sensitivity implied by the vicinity of the cut. One usually invokes what is
called the assumption of either global or local duality to justify the
neglect of this problem \cite{Neub}. Global duality, the weaker form of the
assumption, applies to the case of the SL width, where the integration over
the lepton spectrum is equivalent to an average over the invariant mass of
the final state hadronic system, thus providing an intrinsic source of
smearing. The success of the improved parton model in inclusive hadronic
$\tau$ decay is an empirical argument in support of global duality (for a
recent confirmation see  ref.~\cite{gn}) even at relatively small energies.
The stronger assumption of local duality is instead necessary for inclusive
non-leptonic (NL) decays, where the dynamics is even more complicated
because of the presence in the basic interaction of two hadronic currents
instead of one as in the SL case. Recently arguments against the validity,
in general, of either form of duality have been given in ref.~\cite{shifm}.

In the present  note we argue that, in  spite of the complexity of the
problem,  the  charm and beauty   data  appear  to  indicate a  simple
phenomenological recipe that  considerably improves the situation.  We find 
that the validity   of the usual   approach for the  SL widths is perhaps 
consistent with  the data. In  particular  the SL widths have been 
determined experimentally for  three charmed hadrons, the $D^+$, the  $D^0$
and the $\Lambda_c$ \cite{pdg}  and,  in spite of the large differences in
the corresponding lifetimes, they are close together, with corrections  that
presumably  could be described by the usual theory. Furthemore, the value of
$\vert V_{cb}\vert$ extracted from the inclusive SL B-meson width is in good
agreement, for a reasonable value of the $b$-quark pole mass,  with the
corresponding determination from $B \rightarrow D^*  {\it l} \nu$
\cite{Neub}. In the usual approach, all  widths are predicted to be
proportional   to  the fifth  power  of  the   quark  mass apart  from
corrections of order $1/m^2$ or smaller. On the contrary we will argue that
for the NL widths the presence of unexpected corrections of order $1/m$ is
strongly indicated  by the data.  Not only  that but we  find that these 
$1/m$ corrections are well described  by the simple ansatz that replaces the
quark  mass  with the decaying  hadron mass  in  the $m^5$ factor in front
of the NL width \cite{martihf95}.

This replacement provides a much better description of the NL widths. We
show that, for beauty, both the problems of the SL branching ratio and of
the difference in the lifetimes of the $\Lambda_b$ baryon and the $B$ mesons
are quantitatively solved. For charm a much better fit to the seven known
lifetimes is obtained in terms of four parameters of reasonable size: one
lifetime, one interference contribution for $D^+$, one for $\Xi^+$ and a
smaller W-exchange term for $D^s$.

In the following  we present our analysis in comparison with the standard
one. We first discuss $b$-decays, then $c$-decays  and finally we present
our conclusions.
 \vskip 12pt 
\noindent {\bf 2. Beauty Decays}

The experimental  value of the average SL width of the  $B$ mesons can  be
obtained from the measured  values of the  average SL  branching ratio and
lifetime \cite{pdg}.   The result  is  in good  agreement with the
theoretical  prediction \cite{bigi,Neub}.  This  statement is based on the 
equality (within   errors)  of the   extracted value   of  $\vert
V_{cb}\vert$  compared  with  its  independent determination from  the
exclusive decay $B \rightarrow D^*  {\it l} \nu$ \cite{iw}.  The value of
$\vert V_{cb}\vert$ is obtained from the inclusive SL decay rate of $B$
mesons using the relation \be \Gamma_{SL} (B) = \Gamma_0 \eta_{QCD}
\left[\left( 1 + {\lambda_1 + 3 \lambda_2\over 2m^2_b}\right) I_0 (x,0,0) -
{6\lambda_2\over m^2_b} (1-x)^4  + O (1/m^3_b)\right]\label{eq:1} \ee

\noindent where $\Gamma_0 = (G^2_F m^5_b/ 192\pi^3) \vert V_{cb} \vert^2$,
$I_0 (x,0,0)$ is a phase space factor \be I_0 (x,0,0) = 1 - 8x + 8x^3 - x^4
-12x^2 \log x \label{eq:2} \ee \noindent with $x = (m_c / m_b)^2$ (all
lepton masses are neglected) and $\eta_{QCD}$ is the perturbative QCD
correction (precisely this correction is only appropriate for the first term
in curly brackets, but the second is quite small, and the completely
factorised form is particularly useful for our purposes). The power
suppressed terms $\lambda_1$ and $\lambda_2$  arise from the kinetic-energy
and the chromo-magnetic dimension 5 operators \cite{bigi}--\cite{buv}. We
have $- \lambda_1/2m_b =\langle B \vert \bar h (i\vec{D})^{2} h \vert B
\rangle/2m_b $,  the average kinetic energy of the heavy quark in the
hadron, while  $\lambda_2$ is related to the mass splitting between vector
and pseudoscalar mesons  $\lambda_2 = (m^2_V - m^2_P)/4$. For $B$-mesons, 
current  estimates give $\lambda_1 \sim - 0.4$  GeV$^2$ \cite{lambda1}; 
$\lambda_2 \sim 0.12$ GeV$^2$ is instead obtained from the experimental
squared mass-difference $m^2_{B^*}-m^2_B$. The value of the QCD correction
$\eta_{QCD}$  is affected by considerable uncertainties \cite{bal}. Another
main source of uncertainty arises from the value of $m_b$, the $b$-quark
pole mass. Perturbatively different definitions of $m_b$ result in a change
of $\eta_{QCD}$. As a consequence, here we prefer to make use of the above
expression to obtain $m_b \eta^{1/5}_{QCD}$ from the experimental value of 
$\Gamma_{SL}(B) = B_{SL}(B)/ \tau_B$ and from  $\vert V_{cb} \vert$ as
derived from exclusive decays  ($\vert V_{cb} \vert = (38.6 \pm 2.6)
10^{-3}$ \cite{Neub}). Using for the average SL branching ratio, $B_{SL} (B)
= (10.77\pm 0.43)$\% \cite{Neub,Skwarnicki}, and for the average $B$ meson
lifetime, $\tau_B = (1.55 \pm 0.02)$ ps  \cite{bede}, we find \be m_b
\eta^{1/5}_{QCD} = 4.9 \pm 0.2 \mbox{  GeV}\label{eq:3} \ee

The main uncertainties arise from the $x$ value (taken between 0.08 and 0.12
as suggested by the relation $x = (1-\Delta m / m_b)^2$ with $\Delta m = m_b
- m_c \sim 3.4$ GeV as found in ref.~\cite{Neub}), the expe\-ri\-mental
error on $\vert V_{cb}\vert$, and, to a lesser extent, from the experimental
error on $\Gamma_{SL}(B)$.  The errors on $\lambda_1$ and $\lambda_2$  are
practically irrelevant. For the indicative value $\eta_{QCD} = 0.8$ (in
ref.~\cite{bal} the authors estimated $\eta_{QCD} =0.77 \pm 0.05$) one
obtains about 5.1 GeV for $m_b$, using  eq.~(\ref{eq:3}).  This value is
somewhat large, although in agreement with the lattice results of
ref.~\cite{cgms}, and  compatible with the results obtained from the QCD sum
rules  \cite{nari} or from the analysis of ref.~\cite{bal}. The value $m_b =
$5.1 GeV corresponds to $m_c\sim 1.7$ GeV (using the quoted value for
$\Delta m$) which is also well consistent with the pole mass result for
$m_c$ derived from ref. \cite{almc}, (see below).

The prediction for the SL width of $\Lambda_b$ mainly differs from 
eq.~(\ref{eq:1}) in that $\lambda_2 = 0$. The kinetic energy term
$\lambda_1$ in principle is also different, but in practice its estimate for 
$B$ and $\Lambda_b$ are identical within errors \cite{bigi,Neub}. Some
presumably small additional difference arises from the neglected $1/m^3$
terms. The vanishing of $\lambda_2$ produces a 3.5\% increase of
$\Gamma_{SL}(\Lambda_b)$ with respect to $\Gamma_{SL}(B)$. 

We now consider the NL widths. It is well known that the  observed ratio of
the $\Lambda_b$ to $B$ lifetimes appears too large to be explained by
corrections of order $1/m^2$ or $1/m^3$. This is confirmed by a number of
recent analyses \cite{rosne,neusa}. Here we show that the assumption that
the NL widths scale as the fifth power of the decaying hadron mass (apart
from corrections of order $1/m^2$ and beyond) gives a very good agreement
with experiment. In fact this assumption leads for the ratio of lifetimes to
the expression  \be {\tau_B\over \tau_{\Lambda_b}} = ({m_{\Lambda_b}\over
m_B})^5 \left[ 1-2.24 B_{SL} (B)\right] + 2.24 B_{SL} (B) +
O(1/m^2)\label{eq:4} \ee   \noindent Here the factor 2.24 arises from taking
the electron, the muon and the tau SL modes in the ratio 1:1:0.24
\cite{Neub}, and the difference in the SL rates of the  $\Lambda_b$ and $B$
has been neglected. From $m_{\Lambda_b} = 5623 \pm 6$~MeV \cite{bede} and
$m_B = 5279 \pm 2$~MeV \cite{pdg}, by using the already quoted value for
$B_{SL}$(B), we obtain \be{\tau_B\over \tau_{\Lambda_b}} = 1.29 \pm
0.05\label{eq:5} \ee where the error is dominated by the uncertainty on the
power suppressed corrections of order $1/m^2$ or $1/m^3$.  In comparison,
from the most recent data we have \cite{bede} $\tau_B = 1.55\pm 0.02$ ps 
for the average  $B$ lifetime and $\tau_{\Lambda_b}  = 1.19  \pm 0.06$ ps, a 
value  that includes the  LEP  data and the  recent preliminary result of
CDF. From these values we find \be ({\tau_B\over \tau_{\Lambda_b}})_{EXP} =
1.30 \pm 0.07 \label{eq:6}\ee in perfect agreement with the above
prediction. Clearly it would be very interesting to measure the SL branching
ratio of the $\Lambda_b$ in order  to check whether the SL width is within a
few percent equal to that of the $B$ mesons. Notice that, by neglecting
terms of $O(1/m^3)$, the standard prediction from the heavy quark effectice
theory gives $\tau_B/\tau_{\Lambda_b}=1.02$ \cite{neusa}. Moreover  it  is
very unlikely that the inclusion of the corrections  of $O(1/m^3)$ is 
sufficient to remove the discrepancy \cite{neusa}.

If we repeat the same exercise by applying eq.~(\ref{eq:4}) to the $B_s$ and
$B$ lifetimes, we find  \be {\tau_B\over \tau_{B_s}} = 1.07 \pm
0.03\label{eq:7} \ee where the value $m_{B_s} =5369.6 \pm 2.3$ MeV  from LEP
and CDF was used \cite{bede}. The error from the power suppressed terms can
now be taken smaller than in eq.~(\ref{eq:5})  because of the much closer
similarity of the two mesons involved. The present value of the $B_s$
lifetime, also including the new preliminary data from CDF is given by
$\tau_{B_s} = 1.49 \pm 0.07$ ps \cite{bede}. For the ratio we then obtain
the experimental value \be ({\tau_B\over \tau_{B_s}})_{EXP} = 1.04 \pm
0.05\label{eq:8} \ee

At present the data are not sufficiently precise to check the assumed
dependence on the hadronic mass, but this test could become significant in a
near future.

We now discuss the problem of $B_{SL}(B)$. As well known, the theoretical
prediction for $B_{SL}(B)$ is somewhat larger than the experimental value 
\cite{altpe}.  A possible explanation of this fact could be a failure of the
improved parton model for the $b \rightarrow c \bar{c} s$ mode due to the
restricted phace space for the final state \cite{fawi}. If the rate for this
mode would be sufficiently larger than the predicted value the corresponding
increase of the NL width could reconcile the value of $B_{SL}(B)$ with the
observed result. The problems with this explanation are, on the one hand,
that the observed average number of charm quarks in the final state of
$b$-decay is lower than required. The present experimental result for the
charm counting is given by $n_c = 1.16 \pm 0.05$ \cite{browd}, while the
required amount would be at least $n_c = 1.3$, see the Erratum of
ref.~\cite{babab}. On the other hand, the same mechanism clearly cannot be
invoked to explain the ratio of the $\Lambda_b$ and the $B$ lifetimes.  At
lowest order in $1/m$, a different,  larger $b \to c \bar c s $ rate would
indeed modify identically the $\Lambda_b$ and $B$ lifetimes. On the contrary
a modest increase of the effective $m^5$ factor in front of the NL channels
with respect to that of the SL width decreases the value of $B_{SL}(B)$ to
the observed value. A recent  accurate analysis in the conventional approach
of $B_{SL} (B)$ leads  \cite{babab} to a predicted value $B^{th}_{SL} (B) =
(12.0 \pm 1.4)$ \%, when the $b$-quark pole mass is used in the $m^5$
factor. If the pole mass is replaced by  the $B$-mass  in the $m^5$ factor,
the central value for $B^{th}_{SL} (B)$ is changed into the new figure
$\widetilde B^{th}_{SL} (B)$  given by
\be \left(\widetilde  B^{th}_{SL}(B)\right)^{-1} =  2.24 + r
\left[\left(B^{th}_{SL} (B)\right)^{-1} - 2.24\right]\label{eq:9} \ee
Inserting $r=(5.279/5.1)^5= 1.188$, from $B^{th}_{SL}(B) = 12\%$ we find 
$\widetilde  B^{th}_{SL}(B)=0.105$. Note that the value 5.1 GeV for the pole
mass, as inferred from the SL width, being on the upper side of the error
band for this quantity, leaves space for a larger adjustement if the
preferred value  of $B^{th}_{SL}(B)$ is larger.  In fact, in the analysis of
ref.~\cite{babab}, the value of $B^{th}_{SL}(B)=12.0 \pm 1.4 \, \%$ given
before  corresponds to $n_c= 1.24 \pm 0.05$, which is still too large with
respect to the experimental value. For example, for $n_c\sim 1.16$ and
$\widetilde B^{th}_{SL}$ equal to  the experimental value $ \widetilde
B_{SL}(B)=10.8\%$ one obtains $B^{th}_{SL}(B)\sim 13\%$ from
ref.~\cite{babab} which leads to $m_b\sim$ 5 GeV.

In conclusion the problems for the inclusive $b$-decay phenomenology seem to
be solved with the replacement of the  quark with the hadron mass in the
$m^5$ factor in front of the NL width. As we shall see this is further
confirmed by the analysis of charm decays.

\newpage  
\noindent {\bf 3. Charm Decays} 
\vskip 12pt Up to
date, seven charmed particle lifetimes have  been  measured  and in three
cases also the SL branching ratio is known, so that the corresponding  SL
width can be extracted. All the available data are collected in table 1. We
start with the simplest case of the SL width. Up to terms of order $1/m^3$,
which could be important but are more difficult to estimate \cite{bloshi},
we have in the conventional theory  (omitting, for simplicity, Cabibbo
suppressed channels) an expression which is completely analogous to
eq.~(\ref{eq:1}), with the obvious replacements of $m_b$, $V_{cb}$,
$x=(m_c/m_b)^2$ with $m_c$, $V_{cs}$, $x=(m_s/m_c)^2$. The value of
$\lambda_2$ vanishes for $\Lambda_c$ (and $\Xi_c$)\cite{bigi,Neub}.  In the
calculation of the inclusive widths of the  $D^+$ and $D^0$ we have used
$\lambda_2 = 0.14$ GeV$^2$ obtained from the experimental value of the
difference $\lambda_2 = (m^2_{D^*} - m^2_D)/4$. For $\lambda_1$ a value
around $-0.4$~GeV$^2$ has been used     for both $D^{0,+}$ and $\Lambda_c$.
The quantity $x$ is very small and we have taken  $I_0 \sim 0.91$.

\begin{table} \centering \begin{tabular} {||c|c|c|c|c|}\hline Hadron    & 
Mass  (MeV/$c^2$) & $ \tau$ (ps)   & $ B_{SL} ($\%$) $ & $ \Gamma_{SL}=
B_{SL}/ \tau$ (ps$^{-1}$) \\ \hline $D^{\pm}      $  & $1869.4 \pm 0.4$ &
$1.057 \pm 0.015$ & $17.2 \pm 1.9$ &$16.3\pm 1.8$\\ $D^{0}        $  &
$1864.6 \pm 0.5$ & $0.415 \pm 0.004$ & $ 8.1 \pm 1.1$ &$19.5\pm 2.6$\\
${D_s}        $  & $1968.5 \pm 0.7$ & $0.467 \pm 0.017$ &                &             
\\ $\Lambda_c^{0}$  & $2285.1 \pm 0.6$ & $0.200 \pm 0.011$ & $ 4.5 \pm 1.7$
&$22.5 \pm 8.5$\\ $\Xi_c^{0}    $  & $2470.3 \pm 1.8$ & $0.098 \pm 0.019$ &               
&              \\ $\Xi_c^{\pm}  $  & $2465.1 \pm 1.6$ & $0.350 \pm 0.055$ &              
&           \\ $\Omega_c^{0} $  & $2704   \pm 4  $ & $0.055 \pm 0.023$ &              
&           \\ \hline \end{tabular} \caption{ Properties of charmed mesons
and baryons; the $\Omega^{0} $ values are our average of the data quoted in
ref.~\protect\cite{wa89}.} \end{table}

For D mesons,  there is a strong  cancellation between the term containing
$\lambda_1 + 3\lambda_2$ and the one,  proportional to $\lambda_2$ in
eq.~(\ref{eq:1}) (note that, in this case, it is appropriate  to restore the
factor $\eta_{QCD}$ at his place in front of $I_0$). This makes the
prediction very unstable, with a central value around $\Gamma_{SL}(D) = 0.29
\, \Gamma_0$ for $\eta_{QCD}=0.7$. Also, the smallness of the coefficient
with respect to unity makes the neglect of the $1/m^3$ terms, which we know
could be large especially for mesons,  totally unjustified. For $\Lambda_c$
the prediction is much more stable, and within a $\pm 10\%$ accuracy, one
finds $\Gamma_{SL}(\Lambda_c) = 0.59 \, \Gamma_0$. The value of $m_c
\eta^{1/5}_{QCD}$ required to reproduce the experimental result for
$\Gamma_{SL}(\Lambda_c)$ is around $m_c \eta^{1/5}_{QCD} = 1.5$~GeV, which
is slightly large but not  unreasonable. For example, by taking the
$\overline{MS}$ charm-quark mass computed in lattice simulations,
$m_c^{\overline{MS}}(\mu=2$ GeV$)=1.48 \pm 0.28$ GeV \cite{almc}, we get for
the pole mass $m_c \sim 1.6$--$1.7$ GeV in agreement with  $m_c
\eta^{1/5}_{QCD} = 1.5$~GeV if we take $\eta_{QCD} =0.7$. In conclusion, the
large uncertainties present for charm and the limited number of  the
existing data  on $\Gamma_{SL}$ prevent  a stringent test of the theory,
which is however consistent with the existing information (given in table
1).

We now consider the lifetimes of charmed particles.  At lowest order in the
$1/m$ expansion,  a much better  agreement with the experimental results 
for the lifetimes is obtained by replacing the heavy-quark mass by the
hadron masses  in the $m^5$ term of  the expression for  $\Gamma_{NL}$. We
neglect at this stage any other mass correction and we write  $\Gamma_{NL}
(m) = \Gamma_{tot}(m) - 2\Gamma_{SL}$, where for $\Gamma_{SL}$ we insert a
universal value chosen as the average of the experimental values for $D^+$,
$D^0$ and $\Lambda_c$, or  $\Gamma_{SL} = (0.174 \pm 0.015) $ps$^{-1}$
\cite{pdg}. The dependence on the hadron mass of $\Gamma_{NL}(m)$ will be
taken according to $\Gamma_{NL}(m) = (m/m_0)^n \Gamma_{NL}(m_0)$ with n = 5,
where $m_0$ is around the average mass of the relevant hadron. We then have
\be \Gamma_{tot} (m) = \tau^{-1} (m) = \tau^{-1} (m_0) ({m\over m_0})^n +
2\Gamma_{SL} (1 - ({m\over m_0})^n)\label{eq:10} \ee

\begin{figure}[t]   
\begin{center}
\setlength{\unitlength}{1truecm} 
\begin{picture}(7.5,10.) 
\put(-8.0,-9.5)
{\includegraphics{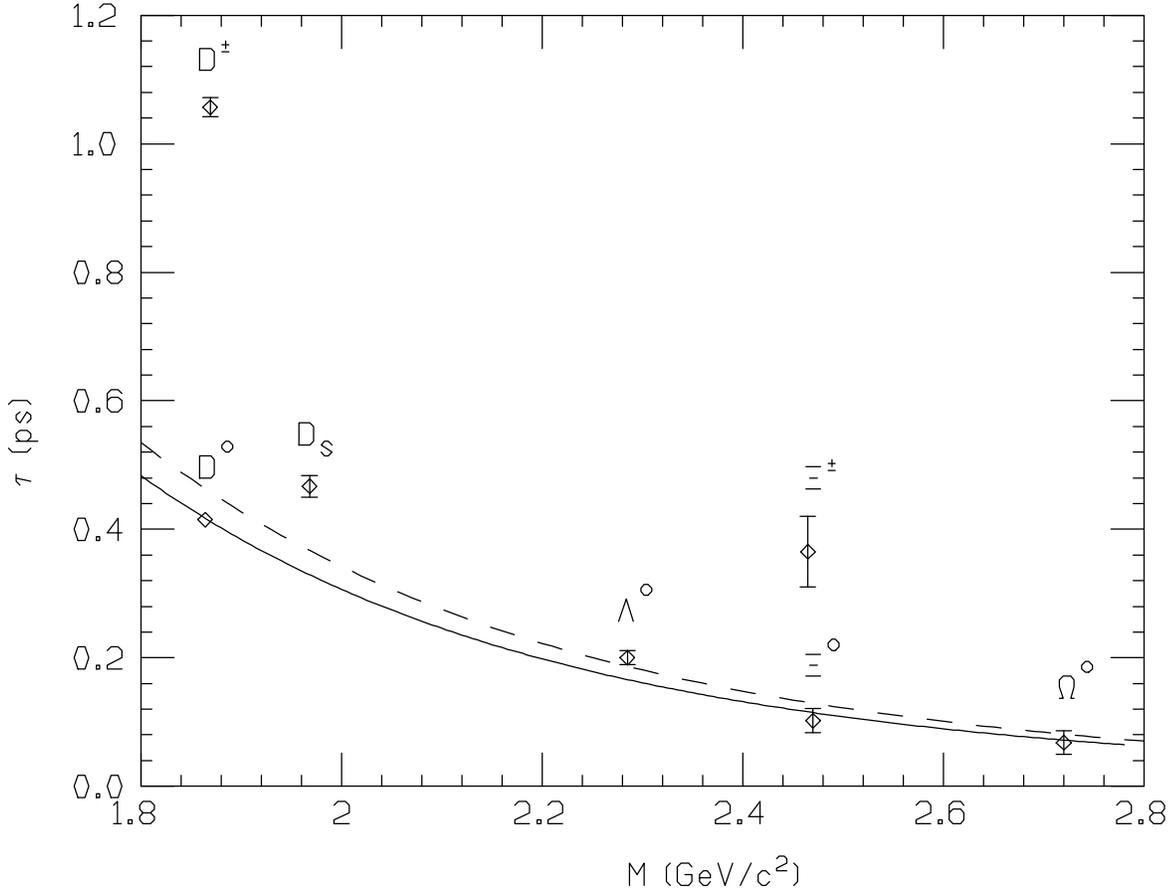}} 
\end{picture} 
\end{center} 
\vskip 1.0cm 
\caption{Lifetime vs. mass for charmed particles. The dashed line is the
best fit described in the text for all seven points assuming proportionality
of the NL widths to $m_{H}^5$, $m_H$ being the hadron mass. The solid line
is the best fit restricted to only the $D^0$, $\Lambda_c$, $\Xi^0$ and
$\Omega_c$ lifetimes in the same assumptions as before.}
\protect\label{Fig.1}
\end{figure}
 
We first fix $n=5, m_0 = 2.3$ GeV and $\Gamma_{SL}= 0.174 \, $ ps$^{-1}$ and
fit all seven known lifetimes in terms of  $\tau (m_0)$. We obtain $\tau
(m_0) = 0.181$ ps. The corresponding fit is shown in fig. 1 (dashed curve).
We see that four out of seven lifetimes are in very good agreement with the
fitted curve. The lifetimes of $D^+$, of $\Xi^+$  and, to a lesser extent,
of $D_s$ are clearly out. We attribute the discrepancies for $D^+$ and
$\Xi^+$ to the interference effect \cite{bigi}. Note that $D^+$  is the only
meson that can have interference at the Cabibbo allowed level and $\Xi^+$ is
the only baryon that can have double interference, in the sense that  $\Xi^+
= cus$ and both $u$ and $s$ can interfere with the corresponding quarks from
$c \rightarrow su \overline {d}$. For $D_s$  the observed smaller difference
is attributed to the possibility of W-exchange \cite{bigi}. All of these
effects are of order $f^2_D /m^2$ or $1/m^3$.  The solid line in fig. 1 has
been obtained  from a modified fit where only  the $D^0$, $\Lambda_c$,
$\Xi^0$ and $\Omega_c$ lifetimes have been considered. In this case, we
obtain  $\tau (m_0) = 0.161$ ps, with the respectable value of the
$\chi^2/d.o.f.$ given by $\sim 3.5$. For  comparison, the fit of the quark
mass to constant lifetimes results  in a  $\chi^2/d.o.f.\sim  251$. 
Finally, for the same four lifetimes, we fit the  power $n$ in
eq.(\ref{eq:10}), keeping fixed the value of  $m_0$  and $\tau(m_0)$ at the
observed values for the $D^0$ meson. In this way we check whether the best
power for $n$  is close to $5$.  We find $n = 4.5 \pm 0.5$, where the error
arises from the experimental errors on the lifetimes. Moreover, if we write
for $D^+$, $\Xi^+$ and $D_s$  the expression $ \tau^{-1} = \tau^{-1} (m_H)
\left[1 - ({\mu\over m_H})^3\right]$, where $\tau$ is the experimental
number given in table 1 and  $\tau(m_H)$ is taken from the previous fit  to
the four remaining lifetimes (with n=5 ), we find $\mu = 1.6, 2.2$ and
$1.3$~GeV  for $D^+, \Xi^+$ and $D_s$ , respectively. We see that the
resulting values of this correction are large, as it is obvious from fig. 1,
but not unreasonable.

\vskip 12pt 
\noindent {\bf 4. Conclusion} 
\vskip 12pt

We have presented a number of experimental facts that, in our  opinion, make
rather clear that  $\Gamma_{NL}$ for charm and beauty decay approximately
scale with the fifth power of hadron masses apart from corrections of order
$1/m^2$ or smaller. These facts are the ratio of the $\Lambda_b$ and $B$ 
lifetimes, the value of $B_{SL}(B)$ and the charm lifetimes. This conclusion
is at variance with the predictions of the short distance operator expansion
approach augmented by the heavy quark effective theory. In fact, according
to this theory, the relevant mass in the rate  should be a universal quark
mass and no corrections of order $1/m$ should be present once this mass is
used. On the contrary the hadron mass differs from the quark mass by
non-universal terms of order $1/m$: $m_H$=$m_q(1+\overline{\Lambda}_H/m_q +
O (1/m^2_q))$. We recall once more that in principle the validity of the
operator expansion in the timelike region, in the vicinity of the physical
cut, is not at all guaranteed \cite{pqw,shifm}. We therefore attribute the
failure of the short distance approach to a violation of the local duality
property that has to be assumed for NL widths. Apparently the conventional
theory for SL widths is not inconsistent with the data. The experimental
evidence for NL widths calls for a reexamination of the underlying
theoretical framework.

\end{document}